\documentclass[pagenumbers,final]{usenixconf}
\usepackage[final]{texstuff}

\usepackage{color}
\usepackage{commands}
  \title{%
   Augmenting Operating Systems With the GPU
  }
\author{%
  Weibin Sun
  \and Robert Ricci%
}

\newcommand{\authoremail}{%
  \ttfamily\char`\{%
  \href{mailto:wbsun@cs.utah.edu}{wbsun},\,%
  \href{mailto:ricci@cs.utah.edu}{ricci}%
  \char`\}@cs.utah.edu%
}

\affiliation{%
  University of Utah, School of Computing \\
  {\normalfont\authoremail\hspace{3ex}\fluxurl}
  \iftechreport
    \\[2ex]
    \textnormal{Flux Technical Note \techreportnumber{}} \\
    \textnormal{\techreportdate{}}%
  \fi
}

\newcommand{\techreportnumber}{FTN--201X--XX}
\newcommand{\techreportdate}{\today}

\date{}

\ifpreprint
  \toappear{%
    Submitted for publication.
  }
\fi

\begin{document}
\maketitle
%% -*- mode: LaTeX -*-
%%

%%%%%%%%%%%%%%%%%%%%%%%%%%%%%%%%%%%%%%%%%%%%%%%%%%%%%%%%%%%%%%%%%%%%%%%%%%%%%%%

%% We can't use `\include', because `\include' creates a page break.  Sigh.
%% -*- mode: LaTeX -*-
%%

%%%%%%%%%%%%%%%%%%%%%%%%%%%%%%%%%%%%%%%%%%%%%%%%%%%%%%%%%%%%%%%%%%%%%%%%%%%%%%%

\begin{abstract}

The most popular heterogeneous many-core platform, the CPU+GPU
    combination, has received relatively little attention in operating
    systems research.
This platform is already widely deployed: GPUs can be found, in some form, in
    most desktop and laptop PCs.
Used for more than just graphics processing, modern GPUs have proved themselves
    versatile enough to be adapted to other applications as well.
Though GPUs have strengths that can be exploited in systems software,
    this remains a largely untapped resource.
We argue that augmenting the OS kernel with GPU computing
    power opens the door to a number of new opportunities.
GPUs can be used to speed up some kernel functions, make other scale better,
    and make it feasible to bring some computation-heavy functionality
    into the kernel.
We present our framework for using the GPU as a co-processor from an OS kernel,
    and demonstrate a prototype in Linux.

%    They have not
%    but have not been applied to systems
%    software.
%So far, however, the notion of integrating GPUs' computing power into operating
%    systems has received only limited attention.

%shown enough
%    versatility to be adopted for other applications as well.
%In this paper, we argue that doing so is both feasible and beneficial:
%we demonstrate that it is possible to effectively make use of GPU
%    computing power in a traditional OS kernel, and that doing so opens
%    new possibilities for performance, functionality, and scaling.
%
%\xxx{Mary doesn't like the word ``extremely'', and thinks we need to
%make the conclusion more clear in the intro.}
%Extremely parallel processors containing hundreds of cores are cheap and
%    widely deployed, in the form of Graphics Processing Units (GPUs).
%Used for more than just graphics processing, modern GPUs have shown enough
%    versatility to be adopted for other applications as well.
%So far, however, the notion of integrating GPUs' computing power into operating
%    systems has received only limited attention.
%In this paper, we argue that doing so is both feasible and beneficial:
%we demonstrate that it is possible to effectively make use of GPU
%    computing power in a traditional OS kernel, and that doing so opens
%    new possibilities for performance, functionality, and scaling.
%We present a prototype that incorporates GPGPU processing into the Linux
%    kernel, and demonstrate its value with a number of early results.
    
\end{abstract}

%%%%%%%%%%%%%%%%%%%%%%%%%%%%%%%%%%%%%%%%%%%%%%%%%%%%%%%%%%%%%%%%%%%%%%%%%%%%%%%

%% End of file.

%% -*- mode: LaTeX -*-
%%

%%%%%%%%%%%%%%%%%%%%%%%%%%%%%%%%%%%%%%%%%%%%%%%%%%%%%%%%%%%%%%%%%%%%%%%%%%%%%%%

\section{Introduction}
\label{sec:introduction}

%\todo{I want to shift the focus of this away from ``neglect'' of the GPU
%    towards the fact that the GPU can have benefits for the OS.}

%\todo{We need to tighten up this section to highlight: (1) GPUs are everywhere,
%(2) What GPUs are good at (3) Preview of the fact that some of the things
%GPUs are good at are useful in OSes (4) Others have looked at heterogeneous
%OSes, but they won't actually work on GPUs (5) Using GPUs from a traditional
%kernel requires a new framework, particularly so that you can run a defined
%set of functions with low latency.}

%The most popular heterogeneous many-core platform, the CPU+GPU
%combination, has received relatively little attention in the OS
%research community.  This platform is already widely deployed: GPUs
%can be found, in some form, in most desktop and laptop PCs.
%Furthermore, this trend is accelerating with the development of
%APUs, %~\cite{APU},
%single chips that combine a CPU with a GPU.  GPUs are
%fairly inexpensive: at the time of writing, a current-generation GPU
%with 336 cores can be purchased for as little as \$160, a price of
%about 50 cents per core.  They are improving at a rapid pace: the
%theoretical performance of the NVIDIA consumer GPUs was improved from
%about 500 gigaFLOPS in 2007 (GeForce 8800) to over 1.3 teraFLOPS in
%2009 (GTX 480)~\cite{CUDA_GUIDE}.

Modern GPUs can be used for more than just graphics processing;
through frameworks like CUDA~\cite{CUDAZONE}, they can run general-purpose
    programs.
While not well-suited to \emph{all} types of programs, they excel on code that
    can make use of their high degree of parallelism.
Most uses of so-called ``General Purpose GPU'' (GPGPU) computation have been
    outside the realm of systems software.
However, recent work on software routers~\cite{packetshader} and encrypted
    network connections~\cite{sslshader} has given examples of how GPGPUs can be
    applied to tasks more traditionally within the realm of operating
    systems.
We claim that these uses are only scratching the surface.
In Section~\ref{sec:applications}, we give more examples of how GPU
    computing resources can be used to improve performance and bring new
    functionality into OS kernels.%
\footnote{In GPU terminology, a program running on the GPU is called
a ``kernel.'' To avoid confusion, we use the term ``OS kernel'' or ``GPU
kernel'' when the meaning could be ambiguous.}
These include tasks that have applications on the desktop, on the server,
and in the datacenter.
%The new functionality includes some tasks that are time-consuming or low-throughput on a CPU,
%but can be significantly improved by GPUs.
%These tasks include antivirus software, program
%analysis and verification, filesystem integrity check, SSL/TLS,
%garbage collection, and many others as discussed in Section~\ref{sec:applications}.
%\todo{Explicitly say something about desktop and server applications}

Consumer GPUs currently contain up to 512 cores~\cite{GTX580}, and are
    fairly inexpensive: at the time of writing, a current-generation GPU with
    336 cores can be purchased for as little as \$160, or about 50 cents per
    core.
GPUs are improving at a rapid pace: the theoretical performance of NVIDIA's
    consumer GPUs improved from 500 gigaFLOPS in 2007 (GeForce 8800)
    to over 1.3 teraFLOPS in 2009 (GTX 480)~\cite{CUDA_GUIDE}.
Furthermore, the development of APUs, which contain a CPU and a GPU on the
    same chip, is likely to drive even wider adoption.
This represents a large amount of computing power, and we argue that
    systems software should not overlook it.
%GPUs are becoming even more common with the development of
%APUs, %~\cite{APU},
%single chips that combine a CPU with a GPU.

Some recent OS designs have tried to embrace processor heterogeneity.
Helios~\cite{Helios} provides a single OS image across multiple
    heterogeneous cores so as to simplify program development.
Barrelfish~\cite{Barrelfish} treats a multicore system as a distributed
    system, with independent OS kernels on each core and
    communication via message-passing.
Both, however, are targeted at CPUs that have support for traditional OS
    requirements, such as virtual memory, interrupts, preemption, controllable
    context switching, and the ability to interact directly
    with I/O devices.
GPUs lack these features, and are thus simply not suited to designs that
    treat them as peers to traditional CPUs.
Instead, they are better suited for use as co-processors.

%\todo{The last paragraph seems better than this one...}
%\com{Section~\ref{sec:design} presents our framework for calling GPU code from
%    OS kernels running on the CPU.
%This framework is designed to run the relatively short tasks used by
%    the OS kernel with a minimum of latency.
%We show results from our prototype implementation.}
%    it pre-loads functions onto the
%    GPU and manages buffers in pinned memory for high performance.
%We present 
%latency-optimized framework for
%    calling GPU code from OS kernels and performance results from our prototype
%    implementation.

%An issue that must be dealt with in order to use GPGPU computing in OS
%    kernels is latency:
%most GPGPU applications are
%throughput-oriented~\cite{Garland:2010:UTA}, with more attention given to
%the rate that relatively long-running GPU kernels can sustain than the speed at
%which shorter tasks can be dispatched.
%In contrast, OS kernel tend to be more concerned with the ability to run
%    relatively short tasks with low latency.
%We address this issue in Section~\ref{sec:design}, presenting a
%    latency-optimized framework for calling GPU code from OS kernels and
%    performance results from our prototype implementation.
%\todo{say something about how it still only applicable to relatively
%long running operations}

%\todo{re-work this paragraph - summary is: GPU can be used to
%  accelerate today's kernels, and we show it. Mention new final
%  section.}

Because of this, we argue that GPUs can be and
should be used to augment OS kernels, but that a heterogeneous OS
cannot simply treat the GPU as a fully functional
CPU with different ISA. The OS kernel needs a new framework if it is
to take advantage of the opportunities presented by GPUs.
To demonstrate the feasibility of this idea, we designed
and prototyped \glinux{}, a framework for calling GPU code from
the Linux kernel.
%that allows the Linux kernel
%to call functions on a GPU.
We describe this framework and the challenges
we faced in designing it in Section~\ref{sec:design}%
%, and present an 
%example of the speedups that it can offer.
%We show the speedups of a GPU
%implementation of AES cryptography kernel module to demonstrate the
%potential performance gains with a GPU-accelerated OS kernel. We will
%discuss this framework and challenges in design and implementation of
%such a system in Section~\ref{sec:design}.

%%%%%%%%%%%%%%%%%%%%%%%%%%%%%%%%%%%%%%%%%%%%%%%%%%%%%%%%%%%%%%%%%%%%%%%%%%%%%%%

%% End of file.

%% -*- mode: LaTeX -*-
%%

%%%%%%%%%%%%%%%%%%%%%%%%%%%%%%%%%%%%%%%%%%%%%%%%%%%%%%%%%%%%%%%%%%%%%%%%%%%%%%%

\section{Applications}
\label{sec:applications}

%\todo{These mostly describe early results, and would take more work to be
%    used in actual kernels.}

We have three motivations for offloading OS kernel tasks to the GPU:
\begin{compactitem}
\item To reduce the \emph{latency} for tasks that run more
    quickly on the GPU than on the CPU
\item To exploit the GPU's parallelism to increase the \emph{throughput} 
    for some types of operations, such as increasing the number of
    clients a server can handle
\item To make feasible incorporation of \emph{new functionality} into
    the OS kernel that runs too slowly on the CPU
\end{compactitem}
%
%The additional hundreds of cores of compute resources can not only
%benefit the traditional functionality of a kernel but also some
%time-consuming or high latency works that can not be done inside the
%current kernel because of large time cost for the kernel aiming at
%quick response, but would make valuable sense if they were done inside
%the kernel. This opens a new direction for OS researchers that many
%complicated but important features would be possible in the OS kernel,
%which means potential gains in security, efficiency, functionality and
%performance of the OS.
%
These open the door for new avenues of research, with the potential for
    gains in security, efficiency, functionality,
    and performance of the OS.
In this section, we describe a set of tasks that have been shown to
    perform well on the CPU, and discuss how they show promise for
    augmenting the operating system.
%We will discuss a set of potential or existing applications
%of the GPU for a OS kernel in the following subsections.

\textbf{Network Packet Processing:}
%\label{sec:network-packet-processing}
%
Recently, the GPU has been demonstrated to show impressive performance
    enhancements for software routing and packet processing.
PacketShader~\cite{packetshader} is capable of fast routing table lookups,
    achieving a rate of close to 40Gbps for both IPv4 and IPv6 forwarding
    and at most 4x speedup over the CPU-only mode using two
    NVIDIA GTX 480 GPUs.
For IPSec, PacketShader gets a 3.5x speedup over the CPU.
Additionally, a GPU-accelerated SSL implementation,
SSLShader~\cite{sslshader} runs four times faster than an equivalent
CPU version.

While PacketShader shows the feasibility of moving part of the
network stack onto GPUs and delivers excellent throughput, it suffers
from a higher round trip latency for each packet when compared
to the CPU-only approach.  This exposes the
weakness of the GPU in a latency-oriented computing model: the
overhead caused by copying data and code into GPU memory and then copying
results back affects the overall response time of a GPU computing task
severely. To implement GPU offloading support, OS kernel
designers must deal with this latency problem. Our \glinux{} prototype
decreases the latency of GPU computing tasks with the
techniques discussed in section~\ref{sec:design}.

Though there are specialized programmable network interfaces which can
    be used for packet processing, the CPU+GPU combination offers a
    compelling alternative: the high level of interest in GPUs, and
    the fact that they are sold as consumer devices drives
    wide deployment, low cost, and substantial investment in improving
    them.

%Although the ideal device to run the network processing code is the
%programmable Network Interface Controller (NIC), the CPU+GPU
%architecture is much more practical than the programmable NIC that has
%a CPU on it because of the wide deployment and the low prices of GPUs.
%The hundreds of cores on a GPU also provide the potential benefit in
%%packet processing throughput, especially for a server with very high
%traffic flows.

%\xxx{Do we have any specific numbers to present that are direct
%  comparisons against packetshader, showing that our latency would be
%  lower?}\com{No, we don't. I implemented the CPU version of packet
%  stealing and forwarding only. The GPU version requires more than I
%  thought before.}

\textbf{In-Kernel Cryptography:}
%\label{sec:cryptography}
%
Cryptography operations accelerated by GPUs have been shown to be
feasible and to get significant speedup over CPU
versions~\cite{Harrison_practicalsymmetric, sslshader}.  OS functionality
making heavy use of cryptography includes IPSec~\cite{packetshader} ,
encrypted filesystems, and content-based data redundancy
reduction of filesystem blocks~\cite{tangwongsan:infocom2010} and 
memory pages~\cite{difference-engine}.
%, memory page
%level or higher abstract data structure level are the potential users
%of the GPU implementation of the in-kernel cryptography library.
Another potential application of the GPU-accelerated cryptography is
trusted computing based on the Trusted Platform Module (TPM).  A TPM
is traditionally hardware, but recent software implementations of the
TPM specification, such as vTPM~\cite{vtpm}, are developed for hypervisors
to provide trusted computing in virtualized environments where
virtual machines cannot access the host TPM directly.
%or the host doesn't have TPMs.
Because TPM operations are cryptography-heavy (such as secure hashing
of executables and memory regions), they can also potentially
be accelerated with GPUs.

The Linux kernel contains a general-purpose cryptography library
used by many of its subsystems. This library can easily be extended
to offload to the GPU.
Our \glinux{} prototype
implements AES on the GPU for the Linux kernel, and we present a
microbenchmark in Section~\ref{sec:aes-cuda} showing that it can outperform
the CPU by as much as 6x for sufficiently large block sizes.
%and the microbenchmark
%shows it can outperform the CPU implementation by as much as more than
%6x when the size of data processed is equal or larger than 8KB.
Due to the parallel nature of the GPU, blocks of data can represent
either large blocks of a single task or a number of smaller
blocks of different tasks.
Thus, the GPU can not only speed up bulk data encryption but also
scale up the number of simultaneous users of the cryptography subsystem,
such as SSL or IPSec sessions with
different clients.
%This
%data size can represent large blocks belonging to a single task, or it
%could represent smaller blocks of data belonging to different tasks,
%such as a number of simultaneous SSL connections maintained in the
%kernel.

\textbf{Pattern Matching Based Tasks:}
%\label{sec:pattern-matching}
%
%Pattern matching on strings is a very basic but heavily used task in
%many system level applications.
The GPU can accelerate regular
expression matching, with speedups of up to 48x reported over CPU
implementations~\cite{reg-gpu}.  A network intrusion detection system
(NIDS) with GPU-accelerated regular expression matching~\cite{reg-gpu} 
demonstrated a 60\% increase in overall packet
processing throughput on fairly old GPU hardware.
Other tasks such as information flow
control inside the OS~\cite{information-flow-control}, virus
detection~\cite{gpu-antivirus} (with two orders of magnitude speedup),
rule-based firewalls, and content-based search in 
filesystems can potentially benefit
from GPU-accelerated pattern matching.

\textbf{In-Kernel Program Analysis:}
%\label{sec:pav}
%
Program analysis is gaining traction as a way to enhance the security and
    robustness of programs and operating systems.
For example, the Singularity OS~\cite{Singularity} relies on
    safe code for process isolation rather than traditional memory protection.
Recent work on EigenCFA has shown that some types of program analysis can be dramatically
    sped up using a GPU~\cite{tarun}.
By re-casting the Control Flow Analysis problem (specifically, 0CFA) in terms
    of matrix operations, which GPUs excel at, EigenCFA is able to see a speed
    up of 72x, nearly two orders of magnitude.
The authors of EigenCFA are working to extend it to pointer analysis as well.
With speedups like this, analysis that was previously too expensive to do
    at load time or execution time becomes more feasible;
it is conceivable that some program analysis could be done as code is
    loaded into the kernel, or executed in some other trusted context.
%\todo{say something about userland too?}

%The existing in-kernel program analysis for program verification such
%as what Singularity~\cite{Singularity} does for security can
%significantly benefit from the GPU acceleration. The recent EigenCFA
%paper~\cite{tarun} shows that the GPU-accelerated static control flow
%analysis can get a \emph{72x} speedup over the optimized CPU
%implementation. In the mean while, traditional program analysis is
%considered to be a time-inefficient task that generally does not fit
%into the latency-oriented kernel although the OS kernel can be the
%ideal trusted computing base (TCB) for security-oriented program
%analysis. But now EigenCFA shows it is practical to analyze the
%program with GPU accceleration inside the kernel for security,
%reliability and performance.

\textbf{Basic Algorithms:}
%\label{sec:basalgo}
%
A number of basic algorithms, which are used in many system-level tasks, have
been shown to achieve varying levels of speedup on GPUs.
These include sort, search~\cite{gpgpu-survey} and graph
analysis~\cite{graph-gpu}.  GPU-accelerated sort and
search fit the functionality of filesystems very well. An interesting
potential use of
GPU-accelerated graph analysis is for in-kernel garbage collection (GC).
GC is usually considered to be time-consuming
because of its graph traversal operation, but a recent patent
application~\cite{GPU-GC} shows it is
possible to do the GC on GPUs, and that it may have better performance than on
CPUs. 
Besides GC for memory objects, filesystems also use GC-like operations to
reorganize blocks, find dead links, and check unreferenced blocks for
consistency. Another example of graph analysis in the kernel is
the Featherstitch~\cite{Featherstitch} system, which exposes
the dependencies among writes in a reliable filesystem. One of the most
expensive parts of Featherstich is analysis of dependencies in its
\emph{patch graph}, a task we believe could be done efficiently
on the GPU.

GPGPU computing is a relatively new field, with the earliest frameworks
    appearing in 2006.
Many of the applications described in this section are, therefore, early
    results, and may see further improvements and broader applicability.
With more and more attention being paid to
this realm, we expect more valuable and interesting
GPU-accelerated in-kernel applications to present themselves in the
future.
%than what we can figure out
%now.
%The GPU speedups for the applications in this section are mostly early results,
%as GPGPU computing is only a few years old. It will take more work to
%integrate them into real OS kernels.
%It
%also needs to be pointed out that the GPU-accelerated OS kernel
%examples we described here are just part of the potential applications
%of GPU in an OS kernel.
%%%%%%%%%%%%%%%%%%%%%%%%%%%%%%%%%%%%%%%%%%%%%%%%%%%%%%%%%%%%%%%%%%%%%%%%%%%%%%%

%% End of file.

% -*- mode: LaTeX -*-
%%

%%%%%%%%%%%%%%%%%%%%%%%%%%%%%%%%%%%%%%%%%%%%%%%%%%%%%%%%%%%%%%%%%%%%%%%%%%%%%%%

\section{GPU Computing For The Linux Kernel}
\label{sec:design}
%\todo{Maybe we should use a different name rather than GLinux}
%As we discussed in section~\ref{sec:introduction},
Because of the functional limitations discussed in
Section~\ref{sec:introduction},
it is impractical to run a fully functional OS kernel on a GPU.
Instead, our \glinux{} framework runs a traditional OS kernel on the
    CPU, and treats the GPU as a co-processor.
We have implemented a prototype of \glinux{} in the Linux kernel, using
    NVIDIA's CUDA framework to run code on the GPU.
%\xxx{latency is not so important anymore.}
%simply by compiling the
%architecture-depedent source of the kernel to a different ISA. The OS
%kernel utilizing the CPU+GPU heterogeneous cores needs a new
%architecture. Because of the special programming model of a GPU as
%explained before, the role of the GPU in the CPU+GPU architecture is a
%co-processor providing extremely parallel computing service. We have
%designed and implemented a GPU computing framework, GLinux, for Linux kernel to
%allow kernel tasks being offloaded onto GPUs so as to either be
%accelerated by massively parallelizm or free CPUs from overload. We
%use CUDA to do GPU computing on NVIDIA's graphics cards.

%Currently it is infeasible to operate GPUs for computing from OS
%kernel because of close-sourced GPU drivers and GPGPU computing
%libraries such as the CUDA's runtime library. So GLinux has a
%userspace helper to allocate memory on GPU devices, invoke the
%execution of a CUDA GPU kernel, transfer data between GPU devices and
%main memory and all CUDA-related operations that should have been but
%can not be done inside the kernel because of the close-sourced drivers
%and runtime libraries. We expect the helper to be eliminated and
%integerated into the OS kernel in future when the GPU specifications
%and the source of drivers and runtime libraries are opened.

%In order to 
%Besides the limited programming model of GPUs, there are still
%some challanges the GLinux needs to handle. We discuss them in the
%following subsections.

\subsection{Challenges}
\label{sec:challenges}

\glinux{} must deal with two key challenges to efficiently use the
    GPU from the OS kernel: the overhead of copying data back and forth, and
    latency-sensitive launching of tasks on the GPU.

\textbf{Data Copy Overhead:}
%\label{sec:overhead-data-copy}
%\todo{maybe move this after launch overhead}
A major overhead in GPGPU computing is caused by the fact that the GPU has
    its own memory, separate from the main memory used by the CPU.
Transfer between the two is done via DMA over the PCIe bus.
  %GPU memory and main memory. Memory on GPU can only be copied from
  %and to main memory via DMA over PCIe.
Applications using the GPU
   must introduce two copies: one to move the input to GPU memory,
  and another to return the result. % to main memory.
The overhead of these copies is proportional to the size of the data.
%  copy the input data required
%  by the computing from main memory to GPU memory, copy the result
%  from GPU memory to main memory.  Different from applications doing
%  HPC or graphics processing, the OS kernel is designed to be
%  time-efficient so that the latency of a computing task is very
%  important. Therefore GLinux must strive to reduce the memory copy
%  overhead though the copies are unavoidable.

%In the CUDA case,
%  according to the offical guide~\cite{CUDA_GUIDE},
There are two kinds of main memory the CUDA driver can use: one is general
  memory (called pageable memory in CUDA), allocated by \texttt{malloc()}.
The other is \emph{pinned} memory, allocated by the CUDA driver and
  \texttt{mmap}-ed into the GPU device.
Pinned memory is much faster than the pageable memory when doing DMA.

In \glinux{}, we use pinned memory for all buffers because of its
    superior performance.
The downside of pinned memory is that it is locked to specific physical
    pages, and cannot be paged out to disk; hence, we must be careful about
    managing our pinned buffers.
This management is described in Subsection~\ref{sec:glinux-arch}.

\textbf{GPU Kernel Launch Overhead:}
Another overhead is caused by
the GPU kernel launch, which introduces DMA transfers of the GPU kernel
code, driver set-up for kernel execution and other device-related
operations.
This sets a lower bound on the time the OS kernel must wait for the GPU code
to complete, so the lower we can make this overhead, the more code can
potentially benefit from GPU acceleration.
This overhead is not high when the GPU kernel execution time or the data
copy overhead dominates the total execution time, as is the case for
most GPGPU computing, which is throughput-oriented~\cite{Garland:2010:UTA}.

%This overhead can be ignored when the GPU kernel execution
%time or the data copy time dominates the total service time.
OS kernel workloads, on the other hand, are likely to be dominated by a
    large number of smaller tasks, and latency of each operation is of
    greater importance.
%However, for OS kernel tasks, latency of operations is of higher importance,
%    and many smaller tasks are likely to dominate.
%it is not quite often that the size of the data
%is large enough and the tasks are expected to be time-efficient.
Though larger tasks can be created by batching many small requests, doing
    so increases the latency for each request.
%data can be batched from multiple requests of the same
%service, it may cost more time to wait for enough requests and the
%latency of a single request increases.
CUDA has a ``stream''~\cite{CUDA_GUIDE} technology that allows kernel execution to
    proceed concurrently with GPU kernel execution and data copy.
By itself, this helps to improve throughput, not latency, but we
    make use of it to communicate between code running on the GPU and CPU.
%allows concurrent kernel
%execution and memory copies and is introduced as a way to improve
%throughput (but we will show the ``concurrent'' feature of streams can
%help the CPU-GPU communicaiton discussed below).
%However, the single
%request latency observed by the requester is not reduced.

%\glinux{} strives to reduce the latency of small-sized reqeust for
%latency-optimized OS kernels.

Instead of launching a new GPU kernel every time the OS wants to call a
    GPU code, we have designed a new GPU kernel execution model,
    which we call the Non-Stop Kernel (NSK).
%    Instead of launching a
%GPU kernel every time when a computing is requested, the GPU kernel is
%launched only once at the very beginning.
The NSK is small, is launched only once, and does not terminate.
To communicate with the NSK, we have implemented a new CPU-GPU message-based
communication method.
It allows messages to be passed between the GPU and main memory
while a GPU kernel is still running.
This is impossible in traditional CUDA programming, in which
the CPU has to explicitly wait for synchronization with
the GPU.%
%to know the completion of previous GPU operations.  
We use pinned memory to pass these messages, and NVIDIA's streaming
    features to asynchronously trigger transfers of the message buffer
    back and forth between CPU and GPU memory.
%that can be copied between GPU and main memory
%    asynchronously, in streams for message passing when the GPU execution
%is still on. 
%
Requests are sent from the CPU to the NSK as messages.
The NSK executes the requested service, which it has pre-loaded into
    the GPU memory.
Similarly, the CPU receives completion notifications from the
    NSK using these messages.
%will execute the requestee
%sent to GPUs as messages and the running kernel will call requested service
%functions upon messages.  The CPU side also get notification via
%messages when the service is done.

We measured the time to launch an empty GPU kernel, transfer a small amout
    of input data to it (4KB), and wait for it to
    return. %, both using the traditional method and with our NSK.
%To get the real service time waited by an OS kernel
%requester,
Though most CUDA benchmarks measure only the execution time on the GPU,
    we measured time on the CPU to capture the entire delay the OS kernel
    will observe.
%We use total time observed by the CPU, rather than measuring only the execution
%    time on the GPU, as is done for most CUDA benchmarks.
%an the general GPU timing
%method in CUDA to measure the latencies of both a NSK-based empty
%computing task and a traditional implementation by varying the size of
%data copied for the task from 4KB to 1MB and find that for small tasks
%with 512, 1024 and 2048 GPU threads.
NSK outperforms the traditional launch method by a factor of 1.3x,
    reducing the base GPU kernel launch time to $16.7\mu{}s$ for
    a kernel with 512 threads, $17.3\mu{}s$ for 1024 threads,
    and $18.3\mu{}s$ for 2048 threads.
While this is much larger than the overhead of calling a function on
    the CPU, as we will show in Section~\ref{sec:aes-cuda}, the speedup
    in execution time can be well worth the cost.

%We expect some
%future GPU-accelerated OS kernel tasks to benefit from the NSK model.

Because of a limitation in CUDA that does not allow a running GPU kernel
to change its number of threads dynamically, NSK switches to a
traditional CUDA kernel launch model when a service requires more
threads on the GPU. This switch will not be necessary in future when
the vendors provide the functionality of dynamically creating new
GPU threads.

\subsection{\glinux{} Architecture}
\label{sec:glinux-arch}
\begin{figure}[]
  \centering \includegraphics[width=1.0\linewidth]{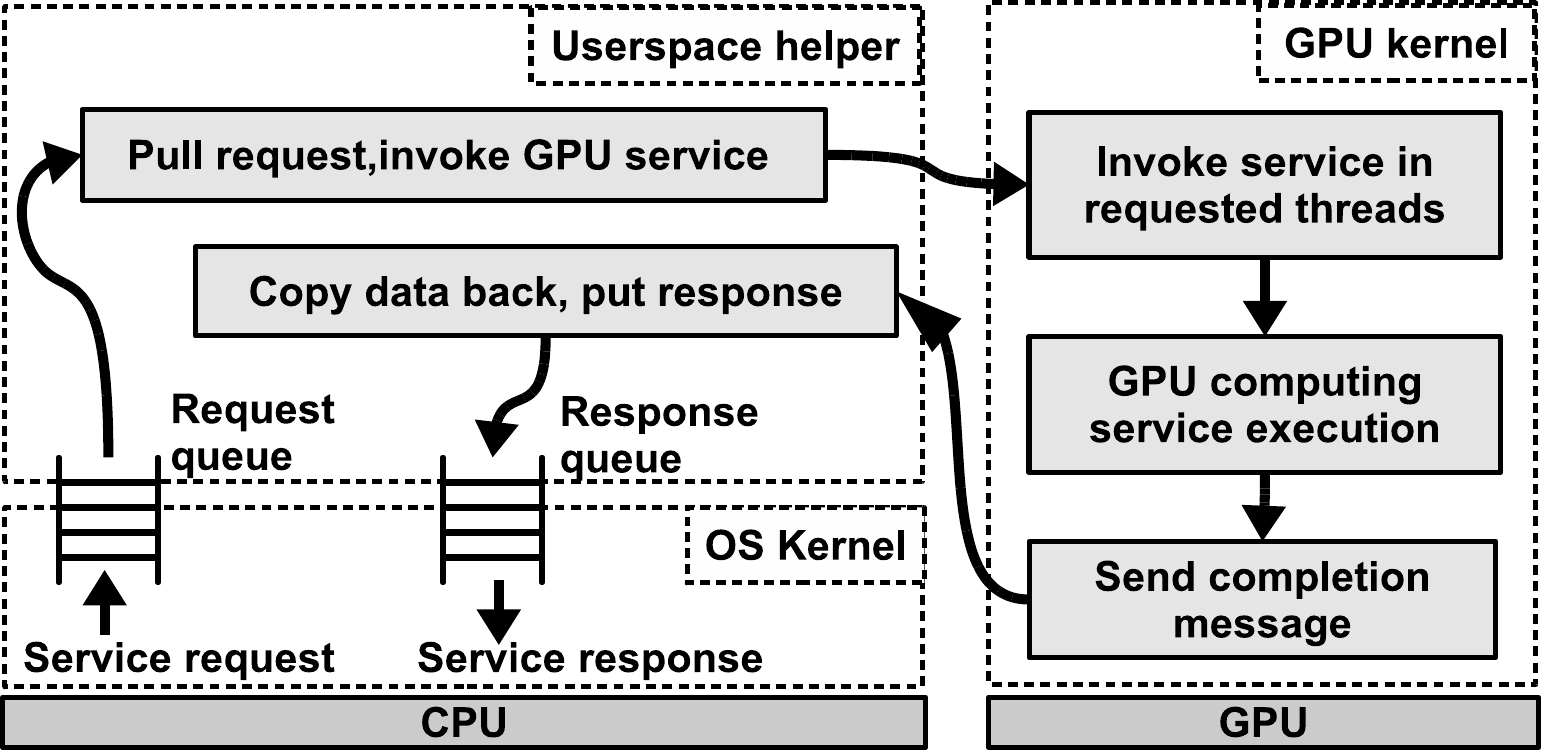}
  \caption{\glinux{} framework architecture}
  \label{fig:arch}
\end{figure}

Our framework for calling the GPU is shown in Figure~\ref{fig:arch}.
It is divided into three parts: a module in the OS kernel, a user-space
    helper process, and NSK running on the GPU.
The user-space helper is necessitated by the closed-source nature of NVIDIA's
    drivers and CUDA runtime, which prevent the use of CUDA directly from
    inside the kernel.
%We believe that this issue is not fundamental, and that if the GPU
%    drivers and libraries were opened, it would be possible move the helper's
%    functionality into the OS kernel.

%To offload a task to the, firstly a kernel
%computing task needs to be implemeted in CUDA. We call a task running
%on GPU a service.  When a running Linux kernel needs to request a
%computing service on GPU, it does the following steps:
%
%It is easy to move an OS kernel computing task onto GPU and to modify
%the code in kernel to use the GPU computing tasks with the help of our
%framework.  The framework as showed in Figure~\ref{fig:arch} works in
%a request-response way.
%
%

To call a function on the GPU, the OS kernel follows the following steps:
\begin{compactitem}
\item It requests one of the pinned-memory buffers, and fills it with the
    input. If necessary, it also requests a buffer for the result.
\item It builds a service request. Services are CUDA programs that have been
    pre-loaded into NSK to minimize launch time. The service request
    can optionally include a completion callback.
\item It places the service request into request queue.
\item It waits for the request to complete, either by blocking until the
    completion callback is called or busy-waiting on the response queue.
%\item when the GPU task completed, the framework sets the response for
%  the request so that the busy waited requester can be noticed, if
%  there is a completion callback, the callback is invoked for the
%  request.
\end{compactitem}

The user-space helper for \glinux{} watches the request queue, which is in
memory shared with the OS kernel.
%checks the queue to dequeue new
%requests.
Upon receipt of a new service request, the helper DMAs the input data buffer
to the GPU using the CUDA APIs. This can proceed concurrently with another
service running on the GPU. When the DMA is complete, the helper sends
a service request message to NSK using the message-passing mechanism
described in Section~\ref{sec:challenges}.
%It then sends a service
%request message to the GPU using the message passing mechanism des
%to invoke the service function on GPU by writing
%to a shared location in memory that is mapped into both the OS kernel and
%the GPU kernel.
When the NSK receives the message, it calls the service function, passing it
pointers to the input buffer and output buffer.
%.  At the GPU
%side, when a service request message is coming, the GPU kernel
%executes the requested service function with the data along the
%request and writes results to the specified output buffer.
When the function completes, the NSK sends a completion message to
the CPU side, and resumes polling for new request messages.
The user-level helper relays the result back to the OS kernel through their
shared response queue.
%The helper receives the completion message and copies the result from
%GPU. Then it puts a response to the response queue to let the
%framework code in Linux kernel notify the service completion to the
%requester as described above.

To avoid a copy between the kernel module and the user-space helper,
the pinned data buffers allocated
by the CUDA driver are shared between the two.
%implement zero-copy.
Also, because NSK allows the user-space helper to work asynchronously
    via messages, service execution on the GPU and data buffer copies
    between main memory and GPU memory can run concurrently.
%    from
%different requests can run concurrently in both the helper and the OS kernel.
As a result, the data buffers locked in physical memory are managed carefully
    to cope with the complex uses.
On the CPU side, buffers can be used for four different purposes:
\begin{compactenum}
\item Preparing for a future service call by accepting data from a caller
    in the OS kernel
\item To DMA input data from main memory to the GPU for the next service call
\item To DMA results from the last service call from GPU memory to main memory
\item Finishing a previous service call by returning data to the caller in the
    OS kernel
\end{compactenum}
Each of these tasks can be performed concurrently, so, along with the service
    currently running on the GPU, the total depth of the service call pipeline
    is five stages.
In the current \glinux{} prototype, we statically allocate four buffers, and
    each changes its purpose over time.
For example, after a buffer is prepared with data from the caller, it becomes
    the host to GPU DMA buffer.
%can be used at the same time for four concurrent requests: two of them are processed 
%in the
%helper by doing the active copying to and copying from the GPU, the other two are processed
%in the kernel space by preparing the next input and doing post-service callback.

On the GPU, we use three buffers: at the same time that one is used by
the active service, a second may receive input for the next service from
main memory via DMA, and a third may be copying the output of the
previous service to main memory.

\subsection{Example: A GPU AES Implementation}
\label{sec:aes-cuda}
To demonstrate the feasibility of \glinux{}, we implemented the AES
encryption algorithm as a service on the GPU for the Linux crypto
subsystem.
%\xxx{Weibin, is it really possible to call it from the
%regular Linux crypto functions?} \com{Yes, at least the kernel module
%itself can do selftest by calling it. I planned to make the eCryptfs
%use this module, but it turned out to be not easy to modify the fs.
%I will push all the benchmark programs into git next week.}
Our implementation is based on an existing CUDA AES
implementation~\cite{engine-cuda}, and uses the ECB cipher mode for
maximum parallelism.
%
%GPU for the Linux kernel crypto
%library basing on an existing CUDA AES
We did a microbenchmark to compare its performance with the original CPU
version in the Linux kernel, which is itself optimized by using special
SSE instructions in the CPU.
We used a 480-core NVIDIA GTX 480 GPU, a quad-core Intel Core i7-930
2.8 GHz CPU and 6GB of DDR3 PC1600 memory. The OS
is Ubuntu 10.04 with Linux kernel 2.6.35.3. 

\begin{figure}[]
  \centering
  \includegraphics[width=1.0\linewidth]{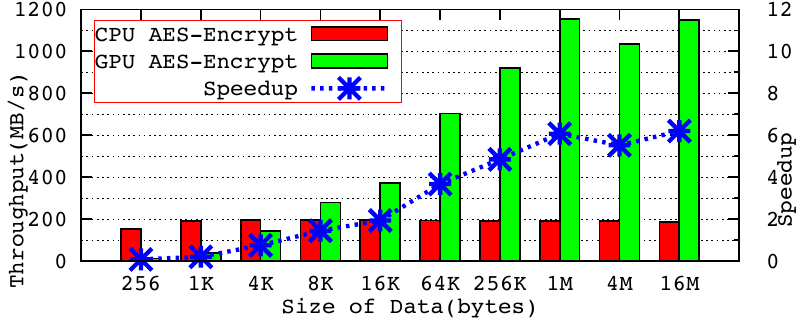}
  \caption{Encryption performance of \glinux{} AES. Decryption, not shown,
  has similar performance.}
  \label{fig:aes}
\end{figure}

We get a performance increase of up to
6x, as shown in Figure~\ref{fig:aes}.
%of AES kernel
%module and got as high as more than 6x speedup.  We listed only the
%encryption performance results in Figure~\ref{fig:aes}.  The
%decryption results are very similar (also up to more than 6x speedup)
%to the encryption one.
The results show that the
GPU AES-ECB outperforms the CPU implementation when the size of the
data is 8KB or larger, which is two memory pages when using typical
page sizes.
So, kernel tasks that depend on
per-page encryption/decryption, such as encrypted
filesystems, can be accelerated on the GPU.
%will get a very good speedup by using the GPU AES-ECB
%computing service.

%This result also shows that not all kernel tasks with
%  all sizes of data can be sped up by GPUs because of the overheads
%  discussed above can't be totally eliminated.
%In the AES case, it is obvious that for AES tasks that have less than 8KB data, the OS
%  kernel should use CPU rather than GPU.
%  This means the OS kernel
%  should be able to decide when to offload the tasks to GPUs basing on
% some hints, like benchmark results in Figure~\ref{fig:aes}. A good
% approach to get such machine-specific results in kernel may be doing
%  some mircobenchmarks for GPU tasks when the OS is booting and then
%  the kernel can decide the offloading policy basing on them.

%%%%%%%%%%%%%%%%%%%%%%%%%%%%%%%%%%%%%%%%%%%%%%%%%%%%%%%%%%%%%%%%%%%%%%%%%%%%%%%

%% End of file.

%% -*- mode: LaTeX -*-
%%

%%%%%%%%%%%%%%%%%%%%%%%%%%%%%%%%%%%%%%%%%%%%%%%%%%%%%%%%%%%%%%%%%%%%%%%%%%%%%%%

\section{Discussion}
\label{sec:discussion}

The GPU-augmented OS kernel opens new opportunities for systems software,
with the potential to bring performance improvements, new functionality,
and security enhancements into the OS.
%researchers with lots of oppotunities which can bring performance improvement, new functionality and security enhancement into OS.

We will continue to develop and improve \glinux{} and to implement more
    GPU functions in our framework.
One such improvement will be dynamically dispatching tasks to the CPU or GPU
    depending on their size.
As seen in Figure~\ref{fig:aes}, the overheads associated with calling the
    GPU mean that small tasks may run faster on the CPU.
Since the crossover point will depend on the task and the machine's specific
    hardware, a good approach may be to calibrate it using microbenchmarks
    at boot time.
% approach to get such machine-specific results in kernel may be doing
%  some mircobenchmarks for GPU tasks when the OS is booting and then
%  the kernel can decide the offloading policy basing on them.
Another improvement will be to allow other kernel subsystems to specifically
    request allocation of memory in the GPU pinned region.
In our current implementation, GPU inputs must be copied into these regions
    and the results copied out, because the pinned memory is used only
    for communication with the GPU.
By dynamically allocating pinned buffers, and allowing users of the framework to
    request memory in this region, they
    can manage structures such as filesystem blocks directly in pinned memory,
    and save an extra copy.
This would also allow multiple calls to be in the preparing and post-service
    callback stages at once.
%An improvement of the buffer management is to provide more buffers
%to allow multiple requests in preparing stage and in post-service callback stage.
%But our prototype uses only four statically allocated, fixed-size buffers; future
%    versions will allocate more buffers on demand.

We expect that future developments in GPUs will alleviate some of the current
    limitations of \glinux{}.
While the closed nature of current GPUs necessitates interacting with
    them from user-space, the trend seems to be towards openness;
AMD has recently opened their high-end 3D GPU drivers and
indicated that drivers for their upcoming APU platform
    will also be open-source.
Furthermore, by combining a GPU and CPU on the same die, APUs, e.g. Intel
SandyBridge and AMD Fusion, are likely
    to remove the memory copy overhead with shared cache between CPU cores and
    GPU cores;
lower copy overhead will mean that the minimum-sized task that can benefit
    from GPU offloading will drop significantly.

{\small
\bibliographystyle{abbrv}
\addcontentsline{toc}{section}{\refname}%
\bibliography{bibliography}
}

%%%%%%%%%%%%%%%%%%%%%%%%%%%%%%%%%%%%%%%%%%%%%%%%%%%%%%%%%%%%%%%%%%%%%%%%%%%%%%%

%% End of file.

\end{document}